\documentclass{statsoc}
\usepackage{url}
\usepackage{epstopdf}
\usepackage{amsmath,amsfonts,natbib,graphicx,psfrag,amssymb}
\usepackage{enumerate,enumitem,verbatim}
\usepackage{color}
\usepackage{mathtools}
\usepackage{geometry}
\usepackage{xcolor}
\usepackage{float}
\pagecolor{white}

\title[A truncated MPCMP model for the analysis of Test match bowlers]{A truncated mean-parameterised Conway-Maxwell-Poisson model for the analysis of Test match bowlers}

\author[P. M. Philipson]{Peter M. Philipson}
\address{
	School of Mathematics, Statistics \& Physics \\
	Newcastle University,
         Newcastle upon Tyne,
         NE1 7RU,
         United Kingdom}
\email{peter.philipson1@newcastle.ac.uk}

\renewcommand{\vec}[1]{\boldsymbol{#1}}

\newcommand{\thetavec}{\vec{\theta}}

\newcommand{\numobs}{45906}
\newcommand{\numtestbowlers}{2148}
\newcommand{\finaltestnum}{2385}
\newcommand{\finaltestdate}{February 2020}

\newcommand{\numyears}{144} 

\begin{document} 

\begin{abstract}
Assessing the relative merits of sportsmen and women whose careers took place far apart in time via a suitable statistical model  is a complex task as any comparison is compromised by fundamental changes to the sport and society and often handicapped by the popularity of inappropriate traditional metrics. In this work we focus on cricket and the ranking of Test match bowlers using bowling data from the first Test in 1877 onwards. A truncated, mean-parameterised Conway-Maxwell-Poisson model is developed to handle the under- and overdispersed nature of the data, which are in the form of small counts, and to extract the innate ability of individual bowlers. Inferences are made using a Bayesian approach by deploying a Markov Chain Monte Carlo algorithm to obtain parameter estimates and confidence intervals. The model offers a good fit and indicates that the commonly used bowling average is a flawed measure.
\end{abstract}


\section{Introduction}
Statistical research in cricket has been somewhat overlooked in the stampede to model football and baseball. Moreover, the research that has been done on cricket has largely focused on batsmen, whether modelling individual, partnership or team scores \citep{Kimber, Scarf2011, Pollard77}, ranking Test batsmen \citep{Brown2009, Rohde2011, Boys2019, Stevenson2021}, predicting match outcomes \citep{Davis2015} or optimising the batting strategy in one day and Twenty20 international cricket \citep{Preston2000, Swartz2006, Perera2016}. Indeed, to the author's knowledge, this is the first work that explicitly focuses on Test match bowlers.

Test cricket is the oldest form of cricket. With a rich and storied history, it is typically held up as being the ultimate challenge of ability, nerve and concentration, hence the origin of the term `Test' to describe the matches. For Test batsmen, their value is almost exclusively measured by how many runs they score and their career batting average, with passing mention made of the rate at which they score, if this is remarkable. Test bowlers are also primarily rated on their (bowling) average, which in order to be on the same scale as the batting average is measured as runs conceded per wicket, rather than the more natural rate of wickets per run.
In this work we consider the problem of comparing Test bowlers across the entire span of Test cricket (1877- ). 

Rather uniquely, the best bowling averages of all-time belong to bowlers who played more than a hundred years ago, contrast this with almost any other modern sport where records are routinely broken by current participants, with their coteries of support staff dedicated to fitness, nutrition and wellbeing along with access to detailed databases highlighting their strengths and weaknesses. The proposed model allows us to question whether the best Test bowlers are truly those who played in the late 19th and early 20th century or whether this is simply a reflection of the sport at the time. Along the way, we also deliberate whether the classic bowling average is the most suitable measure of career performance. Taking these two aspects together suggests that there are more suitable alternatives than simply ranking all players over time based on their bowling average, as seen at 
\url{https://stats.espncricinfo.com/ci/content/records/283256.html}.

The structure of the paper is as follows. The data are described in
Section~2 and contain truncated, small counts which, at the player level, are both under and overdispersed, leading to the statistical model in Section~3. 
Section~4 details the prior distribution alongside the computational details. Section 5 presents some of the results and the paper
concludes with some discussion and avenues for future work in Section~6.

\section{The data}
The data used in this paper consists of $N = \numobs$ individual innings bowling figures by $n=\numtestbowlers$ Test match bowlers from the first Test played in 1877 up to Test \finaltestnum, in \finaltestdate. There are currently twelve Test playing countries and far more Test matches are played today than at the genesis of Test cricket \citep{Boys2019}. World Series Cricket matches are not included in the dataset since these matches are not considered official Test matches by the International Cricket Council (ICC). 

\begin{figure}
    \centering
\includegraphics[height = 3cm]{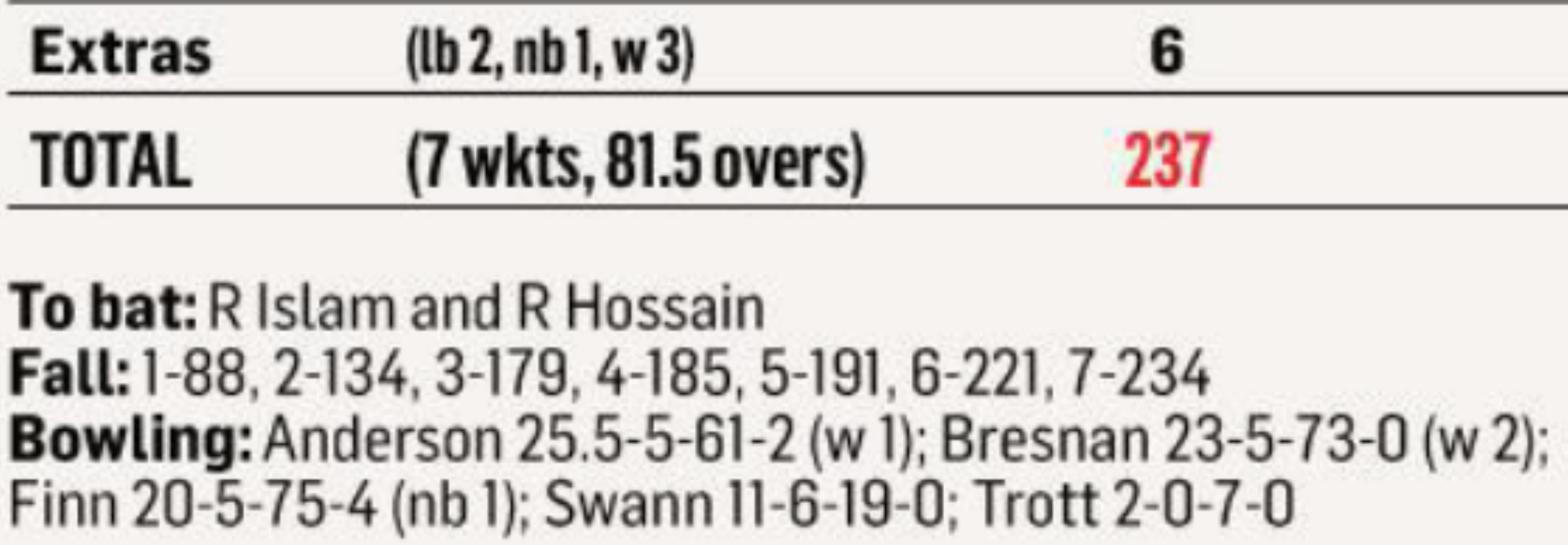}
\caption{Bowling figures as seen on a typical cricket scorecard}
\label{bowlingfigs}
\end{figure}

Bowling data for a player on a cricket scorecard comes in the form `overs-maidens-wickets-runs' - note that in this work the first two values provide meta-information that are not used for analysis -  as seen at the bottom of Figure~\ref{bowlingfigs}. A concrete example are the bowling data for James Anderson with figures of 25.5-5-61-2. The main aspect of this to note is that the data are aggregated counts for bowlers - 2 wickets were taken for 61 runs in this instance, but it is not known how many runs were conceded for each individual wicket. This aggregation is compounded across all matches to give a career bowling average for a particular player, corresponding to the average number of runs they concede per wicket taken. This measure is of a form that is understandable to fans, but counter-intuitive from the standpoint of a statistical model; this point is revisited in subsection~\ref{subsec:runs}.

In each Test match innings there are a maximum of ten wickets that can be taken by the bowling team, and these wickets are typically shared amongst the team's bowlers, of which there are nominally four or five. Table~\ref{Table:TestMatchWickets} shows the distribution of wickets taken in an innings across all bowlers. Taking six wickets or more in an innings is rare and the most common outcome is that of no wickets taken. 

\begin{table}
\caption{\label{Table:TestMatchWickets}Frequency and percentage of Test match wickets per innings}
\centering
\fbox{%
\begin{tabular}{l | c c c c c c c c c}
\hline
Wickets&  0& 1& 2& 3& 4& 5& 6& 7& $\geq$ 8  \\
\hline
Frequency& 9698& 7511& 5726& 3993& 2511& 1557&  637&  245&  80 \\
Percentage& 30.3& 23.5& 17.9& 12.5& 7.9& 4.9& 2.0& 0.8& 0.2 \\
\hline
\end{tabular}}
\end{table}


Alongside the wickets and runs, there are data available for the identity of the player, the opposition, the venue (home or away), the match innings, the winners of the toss and the date the match took place. These are all considered as covariates in the model. 

To motivate the model introduced in section~\ref{sec:model}, the raw indices of dispersion at the player level show that around $25\%$ of players have underdispersed counts, approximately $10\%$ of players have equidispersed counts, with the remainder having overdispersed counts. This overlooks the effects of covariates but suggests that any candidate distribution ought to be capable of handling both under- \textit{and} overdispersion, or \textit{bidispersion}.


\section{The model}\label{sec:model}
Wickets taken in an innings are counts and a typical, natural starting point may be 
to consider modelling them via the Poisson or negative binomial distributions. However, in light of the bidispersion seen at the player level in the raw data, we instead turn to a distribution capable of handling such data.

The Conway-Maxwell-Poisson (CMP) distribution \citep{ConwayMaxwell} is a generalisation of the Poisson distribution, which includes an extra parameter to account for possible over- and underdispersion. Despite being introduced almost sixty years ago to tackle a queueing problem, it has little footprint in the statistical literature, although it has gained some traction in the last fifteen years or so with applications to household consumer purchasing traits \citep{Boatwright2003}, retail sales, lengths of Hungarian words \citep{Shmueli2005},  and road traffic accident data \citep{Lord2008, Lord2010}. Its wider applicability was demonstrated by \cite{Guikema2008} and \cite{Sellers2010}, who recast the CMP distribution in the generalised linear modelling framework for both Bayesian and frequentist settings respectively, and through the development of an R package \citep{COMPoissonReg} in the case of the latter, to help facilitate routine use. 

In the context considered here, the CMP distribution is particularly appealing as it allows for both over- and underdispersion for individual players.
For the cricket bowling data, we define $X_{ijk}$ to represent the number of wickets taken by player $i$ in his $j$th year during his $k$th bowling performance of that year. 
Also $n_i$ and $n_{ij}$ denote, respectively, the number of years in the
career of player $i$ and the number of innings bowled in during year $j$
in the career of player $i$. Using this notation, the CMP distribution has probability mass function given by
\[
\Pr(X_{ijk} = x_{ijk} |\lambda_{ijk}, \nu) = \frac{\lambda_{ijk}^{x_{ijk}}}{(x_{ijk}!)^{\nu}} \frac{1}{G_\infty(\lambda_{ijk}, \nu)}
\]
with $i=1,\ldots, \numtestbowlers, \; j=1,\ldots,n_i$ and  $k=1,\ldots, n_{ij}$. In this (standard) formulation, $\lambda_{ijk}$ is  the rate parameter and $\nu$ models the dispersion. The normalising constant term $G_{\infty}(\lambda_{ijk}, \nu) = \sum_{r=0}^{\infty} \lambda_{ijk}^r/(r!)^{\nu}$ ensures that the CMP distribution is proper, but complicates analysis.

\citet{COMPoissonReg} implemented the CMP model using a closed-form approximation \citep{Shmueli2005, Gillispie2015} when $\lambda_{ijk}$ is large and $\nu$ is small, otherwise truncating the infinite sum to ensure a pre-specified level of accuracy is met.
Alternative methods to circumvent intractability for the standard CMP model in the Bayesian setting have been proposed by \cite{Chanialidis2018}, who used rejection sampling based on a piecewise enveloping distribution and more recently \cite{Benson2020} developed a faster method using a single, simple envelope distribution, but adapting these methods to the mean-parameterised CMP (MPCMP) distribution introduced below is a non-trivial task. 

\subsection{Truncated mean-parameterised CMP distribution}
The standard CMP model is not parameterised through its mean, however, restricting its wider applicability in regression settings since this renders effects hard to quantify, other than as a general increase or decrease. To counter this, two alternative parameterisations via the mean have been developed \citep{Huang2017, Ribeiro2018, Huang2019}, each with associated R packages \citep{mpcmp, cmpreg}. The mean of the standard CMP distribution can be found as
\[
\mu_{ijk} = \sum_{r=0}^{\infty} \frac{r \lambda_{ijk}^r}{(r!)^{\nu}G_\infty(\lambda_{ijk}, \nu)},
\]
which, upon rearranging, leads to
\begin{equation}\label{eq:lam_nonlin}
\sum_{r=0}^{\infty} (r - \mu_{ijk})  \frac{\lambda_{ijk}^r}{(r!)^{\nu}}= 0.
\end{equation}
Hence, the CMP distribution can be mean-parameterised to allow a more conventional count regression interpretation, where
$\lambda_{ijk}$ is a nonlinear function of $\mu_{ijk}$ and $\nu$ under this reparameterisation. \cite{Huang2017} suggested a hybrid bisection and Newton-Raphson approach to find $\lambda_{ijk}$ and applied this in small sample Bayesian settings \citep{Huang2019}, whereas \cite{Ribeiro2018} used an asymptotic approximation of $G_\infty(\lambda_{ijk}, \nu)$ to obtain a closed form estimate for $\lambda_{ijk}$. The appeal of the former is its more exact nature, but this comes at considerable computational cost in the scenario considered here as the iterative approach would be required at each MCMC iteration, and, in this case, for a large number of (conditional) mean values. The approximation used by the latter is conceptually appealing due to its simplicity and computational efficiency, but is likely to be inaccurate for some of the combinations of $\mu_{ijk}, \nu$ encountered here, and the level of accuracy will also vary across these combinations. 

Irrespective of mean parameterisation and method, the above model formulation has two obvious flaws: the counts lie on a restricted range, i.e. {0, \ldots, 10}, and there is no account taken of how many runs the bowler conceded in order to take their wickets.
For the first issue,  the model can be easily modified using truncation, which, in this case, leads to a simplified form for the CMP distribution, which is exploited below:
\begin{eqnarray*}
\Pr(X_{ijk} = x_{ijk} |\lambda_{ijk}, \nu) &=& \frac{\lambda_{ijk}^{x_{ijk}}}{(x_{ijk}!)^{\nu}} \frac{1}{G_{\infty}(\lambda_{ijk}, \nu)} \frac{1}{\Pr(X_{ijk} \leq 10 |\lambda_{ijk}, \nu)} \\
&=& \frac{\lambda_{ijk}^{x_{ijk}}}{(x_{ijk}!)^{\nu}} \frac{1}{G_{10}(\lambda_{ijk}, \nu),} 
\end{eqnarray*}
where $G_{10} = \sum_{r=0}^{10} \lambda_{ijk}^r/(r!)^{\nu}$
 is used to denote the finite sum. This yields the truncated mean-parameterised CMP distribution (MPCMP$_{10}$), where the subscript denotes the value at which the truncation occurs. Furthermore, the infinite sum in (\ref{eq:lam_nonlin}) is replaced by the finite sum
\begin{equation}\label{eq:lam_nonlin_10}
\sum_{r=0}^{10} (r - \mu_{ijk})  \frac{\lambda_{ijk}^r}{(r!)^{\nu}}= 0.
\end{equation}
Since $\lambda_{ijk}$ is positive, there is a single sign change in (\ref{eq:lam_nonlin_10}) when $\mu_{ijk} > r$, which, by Descartes' rule of signs, informs us that there is a solitary positive real root. Hence, a solution for $\lambda_{ijk}$ can be found without recourse to approximations or less scaleable iterative methods in this case and we can directly solve the tenth order polynomial using the \verb|R| function \verb|polyroot|, which makes use of the Jenkins-Traub algorithm. The substantive question of the relationship between wickets and runs is deliberated in the next subsection.


\subsubsection{Functional form of runs}\label{subsec:runs}
As noted earlier, data are only available in an aggregated form. 
That is, the total number of wickets is recorded alongside the total number of runs conceded. Historically, this has been converted to a cricket bowling average by taking the rate of runs conceded to wickets taken, chiefly to map this on to a similar scale as to the classic batting average. 
However, in the usual (and statistical) view of a rate this is more naturally expressed as wickets per run, rather than runs per wicket, and this rate formulation is adopted henceforth. In either event, the number of runs conceded conveys important information since taking three wickets at the cost of thirty runs is very different to taking the same number of wickets for, say, ninety runs. 

\begin{figure}
    \centering
    \includegraphics[width=0.44\textwidth]{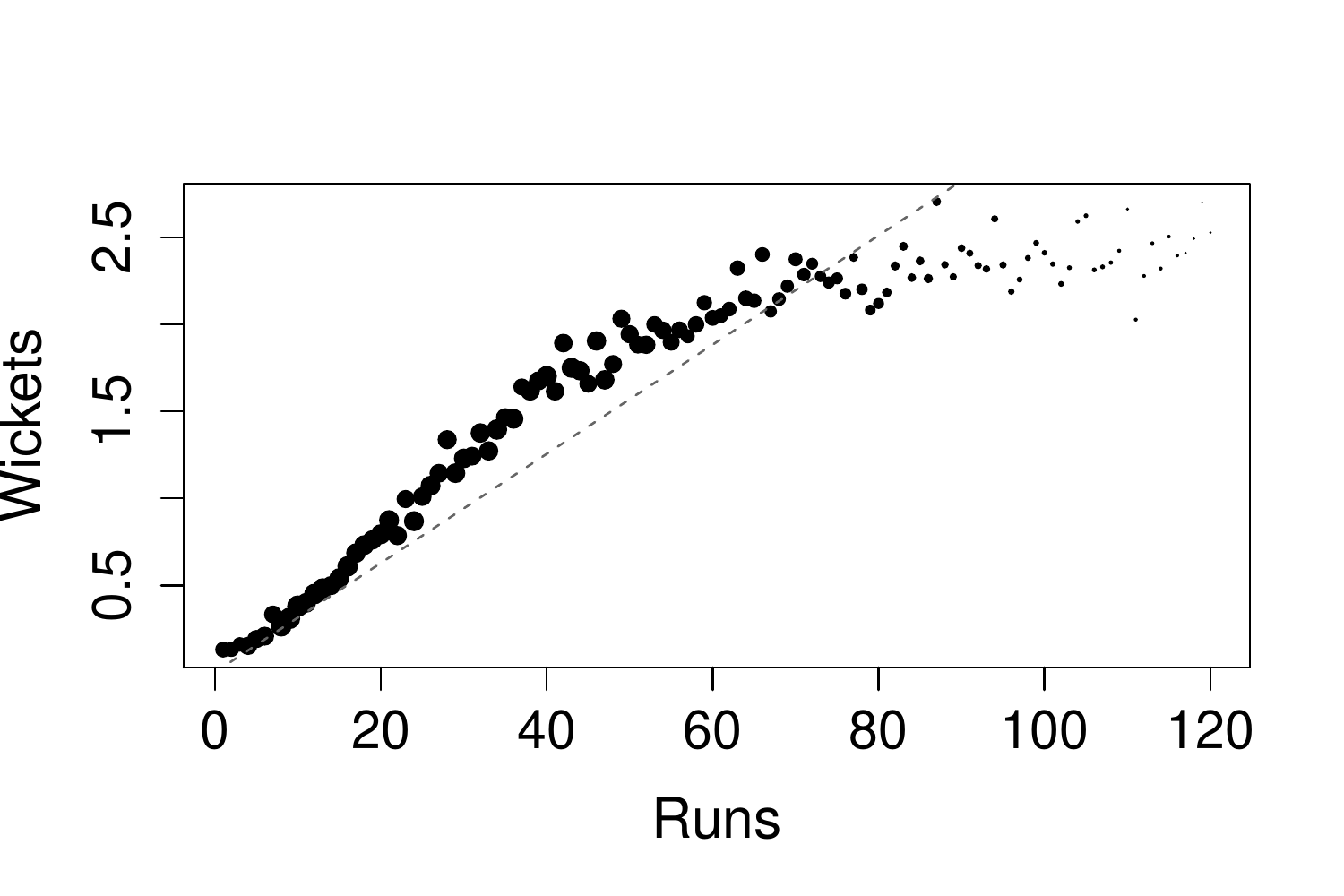}
     \includegraphics[width=0.44\textwidth]{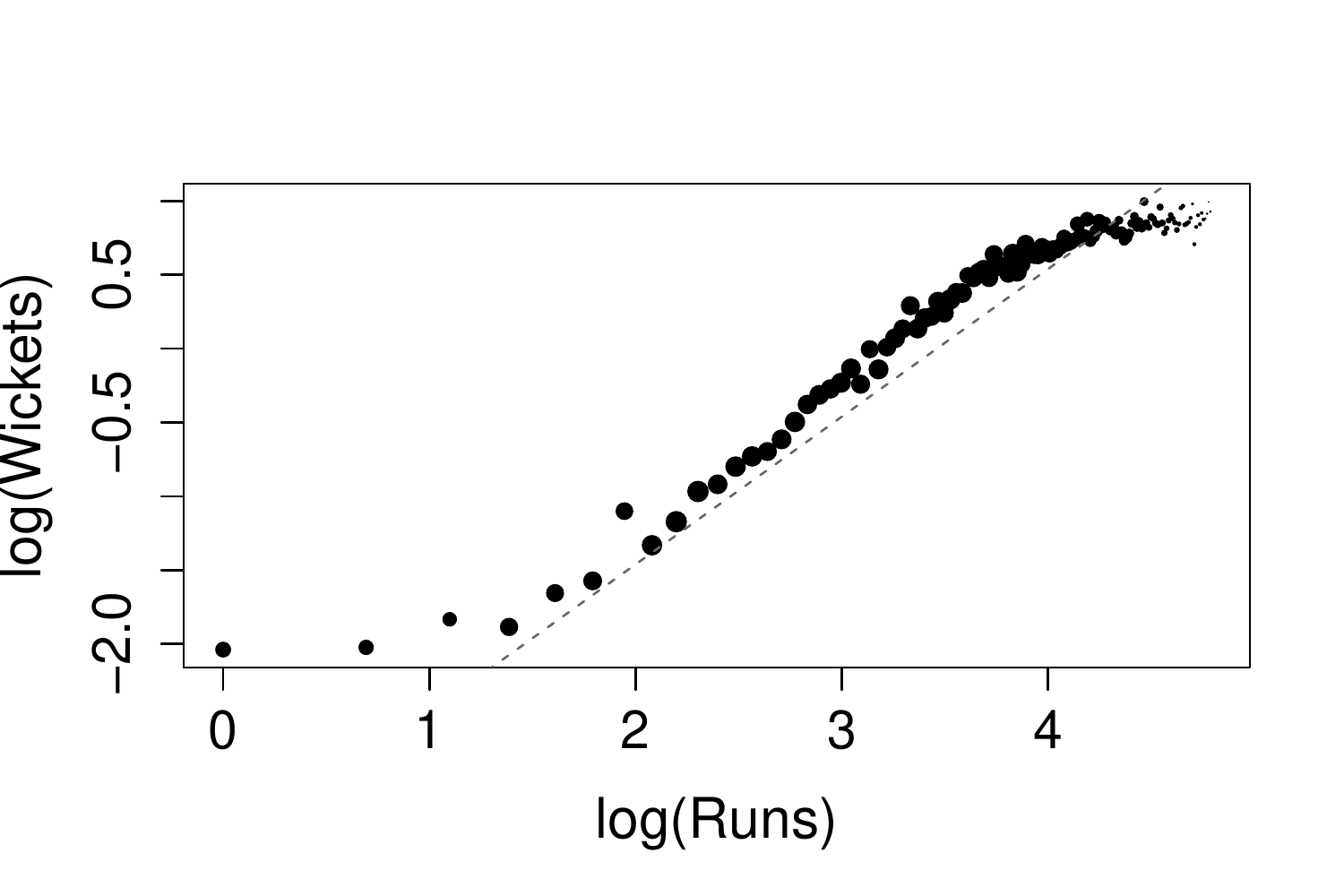}
    \caption{Average wickets taken for values of runs on the linear (left) and log (right) scales. The size of the data points reflects the amount of data for each value of runs; the dashed line represents the relationship under the traditional cricket bowling average.}
    \label{Fig:wkts_vs_runs}
\end{figure}


By looking at the mean number of wickets for each value of runs the relationship between wickets and runs can be assessed, see Figure~\ref{Fig:wkts_vs_runs}; note that there is very little data for values of runs exceeding 120 so the plot is truncated at this point. The nonlinear relationship rules out instinctive choices such as an offset or additive relationship - this makes sense as the number of wickets taken by a bowler cannot exceed ten in a (within-match) innings, which suggests that the effect of runs on wickets is unlikely to be wholly multiplicative (or additive on the log-scale).


Smoothing splines in the form of cubic B-splines are adopted to capture the nonlinear relationship, with knots chosen at the quintiles of runs (on the log scale). This corresponds to internal knots at 18, 35, 53 and 77 along with the boundary knots at 1 and 298 on the runs scale. Various nonlinear models were also considered but did not capture the relationship as well as the proposed spline.

\subsubsection{Opposition effects}\label{subsec:opp_data}
\begin{figure}
    \centering
    \includegraphics[width=0.6\textwidth]{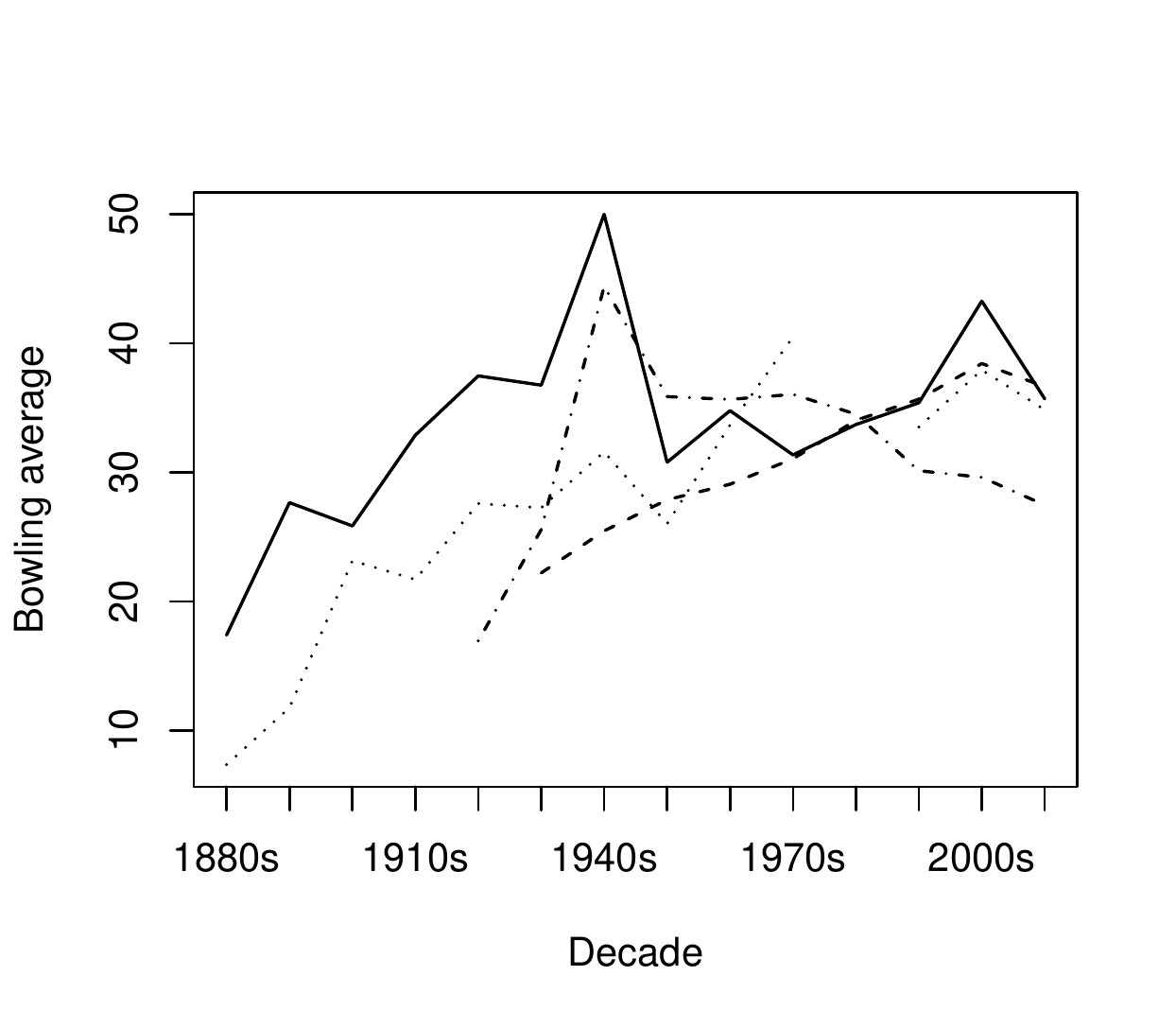}
\caption{Mean bowling averages across decades for Australia (solid line), India (dashed), South Africa (dotted) and West Indies (dot-dash).}
\label{RawOppPlot}
\end{figure}

As the data span $\numyears$ years it is perhaps unreasonable to assume that some effects are constant over time. In particular, teams are likely to have had periods of strength and/or weakness whereas conditions, game focus, advances in equipment and technology may have drastically altered playing conditions for all teams at various points in the Test cricket timeline. A plot of the mean bowling average across decades for a selection of opposition countries is given in Figure~\ref{RawOppPlot}. Here we use the conventional bowling average on the y-axis for ease of interpreation, but the story is similar when using the rate. Clearly, the averages vary substantially over time as countries go through periods of strength and weakness and game conditions and rules evolve. As such, treating them as fixed effects does not seem appropriate. Note also that some countries started playing Test cricket much later than 1877 (see the lines for West Indies and India in Figure~\ref{RawOppPlot}) and teams appear to take several years to adapt and improve.

This motivates the inclusion of dynamic opposition effects, where we again adopt smoothing splines, this time with knots chosen at the midpoints of decades (of which there are fourteen); this is a natural timespan, both in a general and sporting sense. For reasons of identifiability, the opposition effect of Australia in 2020 is chosen as the reference value. 
\subsection{Log-linear model for the mean rate}
As well as runs conceded by a bowler and the strength of opposition, there are several other factors that can affect performance, some of which are considered formally in the model. Introducing some notation for available information: $r_{ijk}$ is the number of runs conceded by bowler $i$ during the the $k$th innings in the $j$th year of his career, $y_{ijk}$ represents the year of this same event, $o_{ijk}$ indicates the opposition (there are twelve Test playing countries in these data), $h_{ijk}$ indicates whether the innings took place in the bowler's home country ($1=\text{home}$, $2=\text{away}$), $m_{ijk}$ is the within-match innings index and $t_{ijk}$ represents winning or losing the toss ($1=\text{won}$, $2=\text{lost}$).

These remaining terms in the model for the wicket-taking rate are assumed to be additive and, hence, the mean log-rate is modelled as
\begin{multline}\label{eq:loglinmodel}
\log\mu_{ijk} = \sum_{p=1}^{8}\beta_{p} B_{p}\left\{\log r_{ijk} \right\} 
+ \sum_{q=1}^{q_o}\omega_{q, o_{ijk}} B_{q, o_{ijk}}(y_{ijk})I(\textrm{Opp} = o_{ijk}) \\
+ \theta_i  
+ \zeta_{h_{ijk}} + \xi_{m_{ijk}} + \gamma I(t_{ijk} = 1) I(m_{ijk} = 1)
\end{multline}
where $\theta_i$ represents the ability of player $i$ and the next three terms in the model are game-specific, allowing home advantage, pitch degradation (via the match innings effects), and whether the toss was won or lost, respectively, to be taken into account. Home advantage is ubiquitous in sport \citep{Pollard2005}, and it is widely believed that  it is easier to bowl as a match progresses to the third and fourth innings due to pitch degradation, and winning the toss allows a team to have `best' use of the playing/weather conditions. The effect of losing the toss is ameliorated as the match progresses so we anticipate that this only effects the first innings of the match, and this effect is captured by $\gamma$.

The design matrices $B_p$ and $B_{q, o_{ijk}}$ are the spline bases for (log) runs and each opposition detailed in the previous two subsections, with associated parameters $\beta_p$ and $\omega_{q, o_{ijk}}$. Note that the summation index for the opposition spline varies across countries owing to their different spans of data - as seen in subsection \ref{subsec:opp_data} - with the total number of parameters for each opposition denoted denoted by $q_o, o = 1, \ldots, 12$.
For identifiability purposes we set $\zeta_1 = \xi_1 = 0$, measuring the impact of playing away via $\zeta_2$ and the innings effects relative to the first innings through $\xi_2, \xi_3$ and $\xi_4$. We impose a sum-to-zero constraint on the player abilities, thus in this model $\exp(\theta_i)$ is the rate of wickets per  innings taken by player $i$ relative to the average player, with the remaining effects fixed.

CMP regression also allows a model for the dispersion. Here, recognising that we may have both under and overdispersion at play, we opt for a player-specific dispersion term $\nu_i, i = 1, \ldots \numtestbowlers$.  Naturally, this could be extended to include covariates, particularly runs, but this is not pursued here since the model is already heavily parameterised (for the mean).

\section{Computational details and choice of priors}
R \citep{rcore} was used for all model fitting, analysis and plotting, with the \verb|splines| package used to generate the basis splines for runs and the temporal opposition effects. The Conway-Maxwell-Poisson distribution is not included in standard Bayesian software such as \verb|rstan| \citep{rstan} and \verb|rjags| \citep{Plummer2004} so bespoke code was written in R to implement the model. The code is available on GitHub at \url{https://github.com/petephilipson/MPCMP_Test_bowlers}.
 
For analysis, four MCMC chains were run in parallel with 1000 warm-up iterations followed by 5000 further iterations. Model fits took approximately eight hours on a standard MacBook Pro. A Metropolis-within-Gibbs algorithm is used in the MCMC scheme, with component-wise updates for all parameters except for those involved in the spline for runs, $\vec{\beta}$. For these parameters a block update was used to circumvent the poor mixing seen when deploying one-at-a-time updates, with the proposal covariance matrix based on the estimated parameter covariance matrix from a frequentist fit using the \verb|mpcmp| package. In order to simulate from the CMP distribution to enable posterior predictive checking, the \verb|COMPoissonReg| \citep{COMPoissonReg} package was used. Plots were generated using \verb|ggplot2| \citep{ggplot2} and highest posterior density intervals were calculated using \verb|coda| \citep{Rcoda}.

\subsection{Prior distribution}
For the innings, playing away and winning the toss effects we use zero mean normal distributions with standard deviation $0.5 \log 2$, reflecting a belief that these effects are likely to be quite small on the multiplicative wicket-taking scale, with effects larger than two-fold increases or decreases deemed unlikely (with a 5\% chance a priori). The same prior distribution is used for the player ability terms, $\thetavec$, reflecting that while heterogeneity is expected we do not expect players to be, say, five and ten times better/worse in terms of rate. The coefficients for the splines in both the runs and opposition components are given standard normal priors. We recognise that smoothing priors could be adopted here, but as we have already considered the choice of knots we do not consider such an approach here.

For the dispersion parameters we work on the log-scale, introducing $\eta_i = \log(\nu_i)$ for $i = 1, \ldots, \numtestbowlers$. We adopt a prior distribution that assumes equidispersion, under which the MPCMP model is equivalent to a Poisson distribution. Due to the counts being small we do not expect the dispersion in either direction to be that extreme, allowing a 5\% chance for $\eta_i$ to be a three-fold change from the a priori mean of equidispersion. Hence, the prior distribution for the log-dispersion is $\eta_i \sim N(0, 0.5 \log 3)$ for $i = 1, \ldots, \numtestbowlers$.

\section{Results}

\subsection{Functional form for runs}
\begin{figure}
    \centering
    \includegraphics[width=0.48\textwidth]{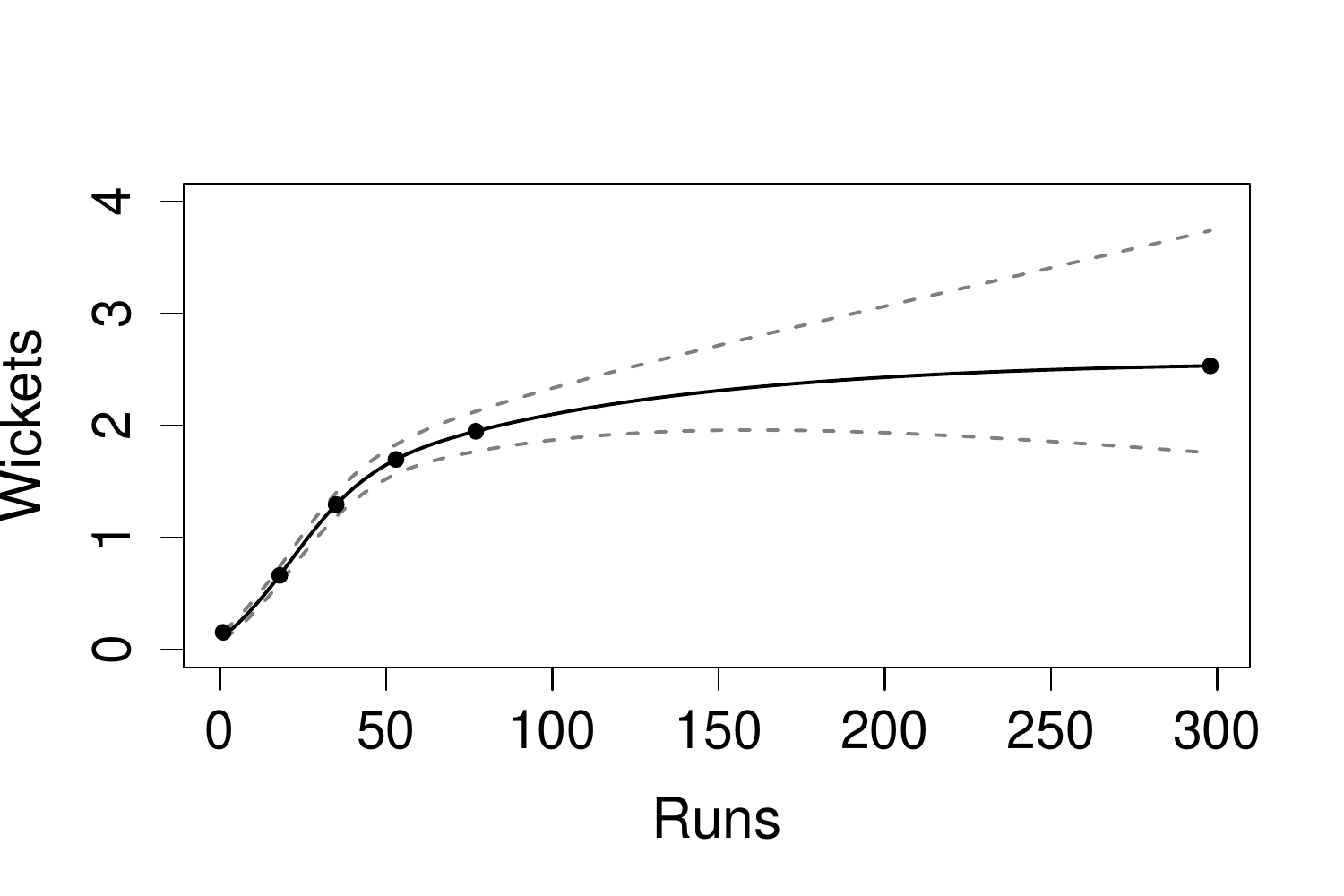}
    \includegraphics[width=0.48\textwidth]{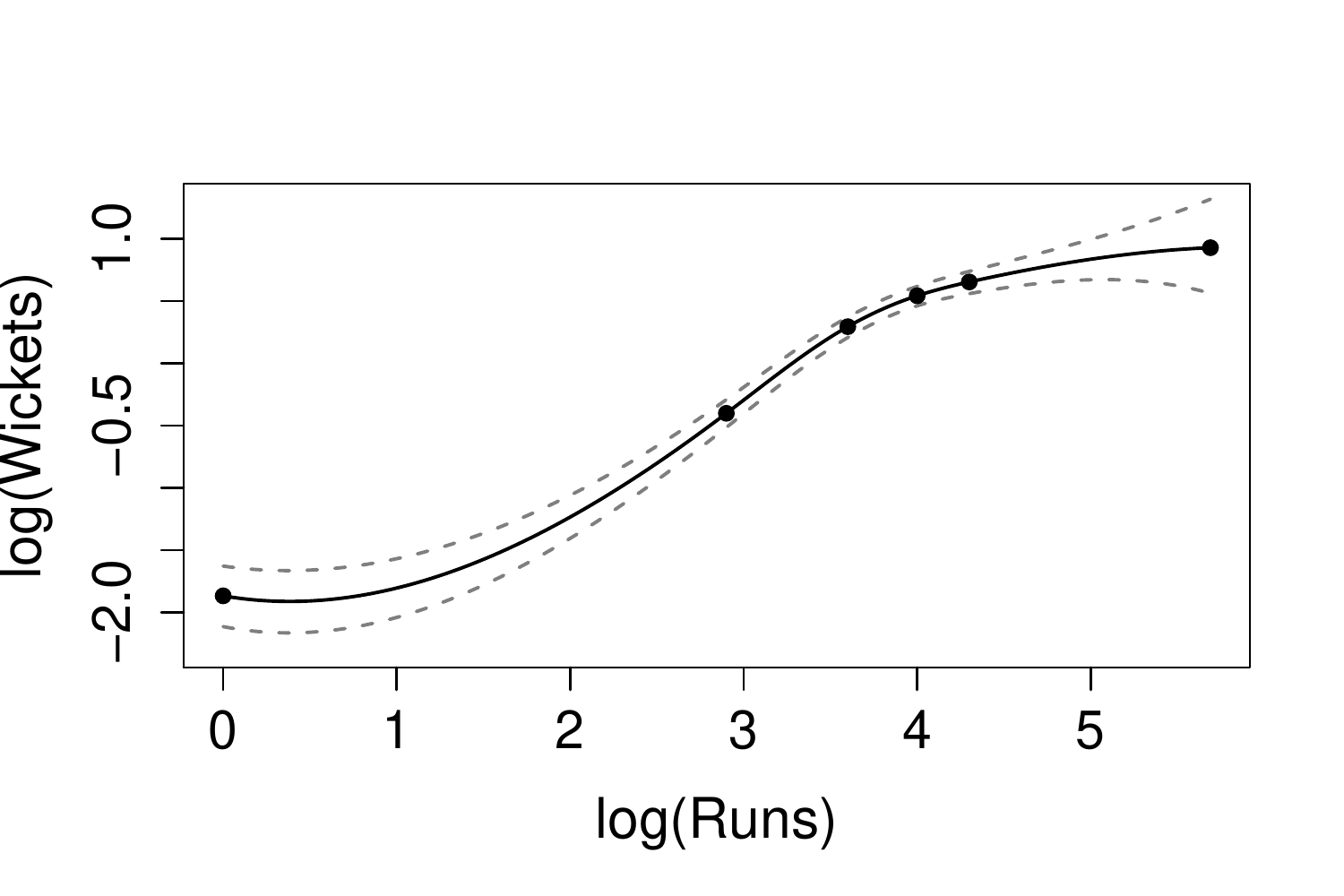}
    \caption{Mean wickets taken against runs on the linear (left) and log (right) scales with 95\% HDI interval uncertainty bands; solid circles represent the knots for the spline}
    \label{fitted_runs_spline}
\end{figure}

A plot of wickets against runs using the posterior means for $\beta$ is given in Figure~\ref{fitted_runs_spline}. This clearly shows the non-linear relationship between wickets and runs and suggests that using the standard bowling average, which operates in a linear fashion, overlooks the true nature of how the number of wickets taken varies with the number of runs conceded. An important ramification of this is that the standard bowling average overestimates the number of wickets taken as the number of runs grows large, whereas the true relationship suggests that the rate starts to flatten out for values of runs larger than 50. Returning to the figure, we clearly see much more uncertainty for larger values of runs, where, as seen earlier, the data are considerably more sparse.

\subsection{Opposition effects}
\begin{figure}
    \centering
    \includegraphics[width=0.85\textwidth]{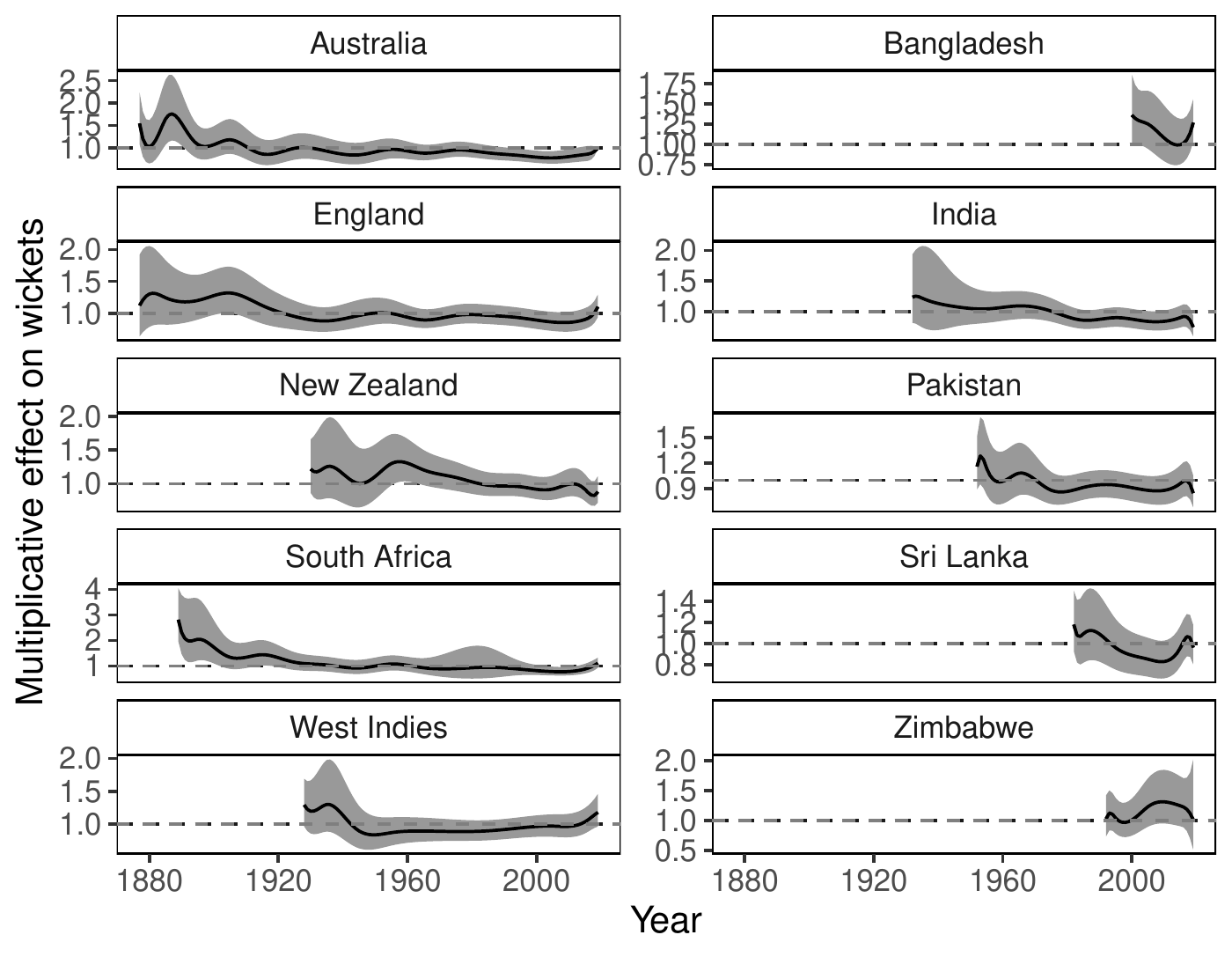}
\caption{Posterior mean and 95\% HDI bands for each opposition} 
\label{fitted_opp_splines}
\end{figure}

A plot of the fitted posterior mean profiles for each opposition is given in Figure~\ref{fitted_opp_splines}. The largest values of all occur for South Africa when they first played (against the more experienced England and Australia exclusively); most teams struggle when they first play Test cricket, as shown by the largest posterior means at the left-hand side of each individual plot. Overall, this led to the low Test bowling averages of the 1880s-1900s that still stand today as the lowest of all-time and adjusting for this seems fundamental to a fairer comparison and/or ranking of players.

\subsection{Game-specific effects}
\begin{figure}
    \centering
    \includegraphics[width=0.45\textwidth]{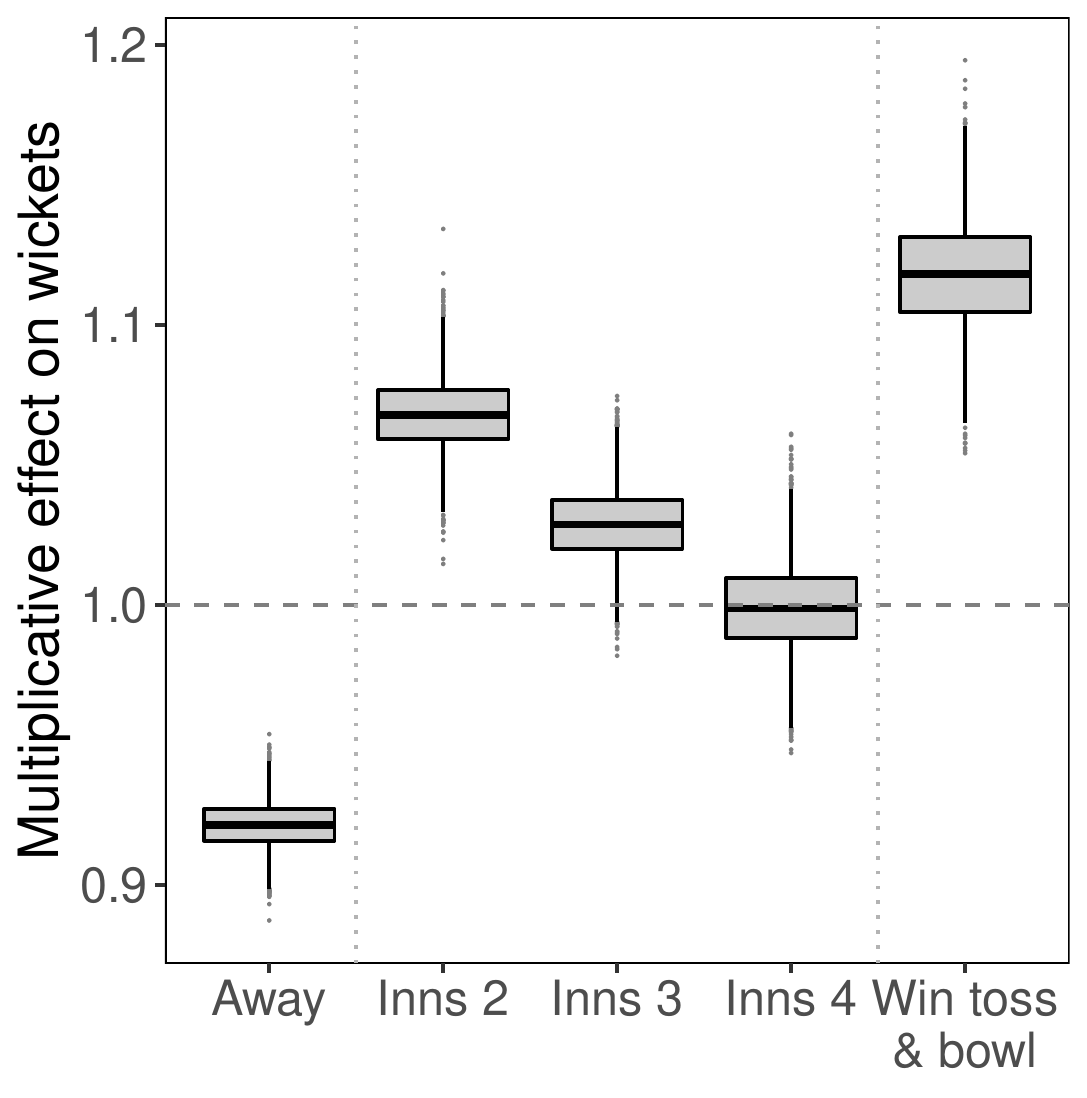}
     \caption{Boxplots of the posterior distributions for playing away ($\zeta_2$), match innings 2-4 ($\xi_2, \xi_3, \xi_4$) and winning the toss and bowling ($\gamma$).}
     \label{FigInnsHA}
\end{figure}

The posterior means and 95\% highest density intervals (HDIs) for the innings effects - on the multiplicative scale - are 1.07 (1.04 - 1.09), 1.03 (1.00 - 1.05) and 1.00 (0.97 - 1.03) for the second, third and fourth innings respectively, when compared to the first innings. Similarly, the effect of playing away is to reduce the mean number of wickets taken by around $10\%$; posterior mean and 95\% HDI are 0.92 (0.90 - 0.94). The impact of bowling after winning the toss is the strongest of the effects we consider, with a posterior mean and 95\% HDI interval of 1.12 (1.08 - 1.15). Boxplots summarising the game-specific effects are shown in Figure~\ref{FigInnsHA}.

\subsection{Player rankings}
The top thirty bowlers, as ranked by their posterior mean ability, $\theta_i$, are given in Table~\ref{TableOfRanks}. We also include the posterior dispersion parameter for each player along with information on the era in which they played (via the date of their debut) and the amount of available data as measured through the total number of innings they played in ($n_i$).

\begin{table}
\caption{\label{TableOfRanks}Top 30 Test match bowlers ranked by posterior mean multiplicative rate}
\fbox{%
\begin{tabular}{r l r r r r c}
\hline
& & & & \multicolumn{2}{c}{Player ability} & Dispersion\\
Rank& Name& Debut& Innings& $E(e^\theta)$& $SD(e^\theta)$ & $E(\nu)$  \\
\hline
1& M Muralitharan& 1992& 230& 2.04& 0.08& 0.83\\ 
2& SF Barnes& 1901& 50& 1.86& 0.18& 0.64\\ 
3& WJ O'Reilly& 1932& 48& 1.81& 0.18& 0.71\\ 
4& Sir RJ Hadlee& 1973& 150& 1.79& 0.11& 0.59\\ 
5& CV Grimmett& 1925& 67& 1.75& 0.14& 0.80\\ 
6& AA Donald& 1992& 129& 1.75& 0.10& 1.04\\ 
7& MJ Procter& 1967& 14& 1.74& 0.26& 1.12\\ 
8& MD Marshall& 1978& 151& 1.72& 0.10& 0.76\\ 
9& DW Steyn& 2004& 171& 1.71& 0.09& 0.71\\ 
10& GD McGrath& 1993& 241& 1.70& 0.09& 0.65\\ 
11& JJ Ferris& 1887& 16& 1.70& 0.23& 1.03\\ 
12& CEL Ambrose& 1988& 179& 1.70& 0.10& 0.69\\ 
13& SK Warne& 1992& 271& 1.70& 0.08& 0.64\\ 
14& T Richardson& 1893& 23& 1.69& 0.21& 0.93\\ 
15& J Cowie& 1937& 13& 1.69& 0.26& 0.89\\ 
16& JC Laker& 1948& 86& 1.68& 0.14& 0.64\\ 
17& Mohammad Asif& 2005& 44& 1.66& 0.17& 0.76\\ 
18& DK Lillee& 1971& 132& 1.66& 0.10& 0.84\\ 
19& K Rabada& 2015& 75& 1.66& 0.12& 1.15\\ 
20& R Ashwin& 2011& 129& 1.64& 0.10& 0.73\\ 
21& Imran Khan& 1971& 160& 1.63& 0.10& 0.71\\ 
22& CTB Turner& 1887& 30& 1.62& 0.20& 0.68\\ 
23& Waqar Younis& 1989& 154& 1.62& 0.09& 0.77\\ 
24& SE Bond& 2001& 32& 1.62& 0.16& 1.22\\ 
25& H Ironmonger& 1928& 27& 1.62& 0.21& 0.57\\ 
26& J Garner& 1977& 111& 1.62& 0.10& 1.23\\ 
27& K Higgs& 1965& 27& 1.61& 0.19& 1.01\\ 
28& FS Trueman& 1952& 126& 1.61& 0.10& 0.85\\ 
29& AK Davidson& 1953& 82& 1.60& 0.14& 0.63\\ 
30& FH Tyson& 1954& 29& 1.60& 0.20& 0.66\\ 
\hline
\end{tabular}}
\end{table}

The rankings under the proposed model differ substantially from a ranking based on bowling average alone. As a case in point, the lowest, i.e. best, bowling average of all time belongs to GA Lohmann, who is ranked 52nd in our model. Similarly, Muttiah Muralitharan, who tops our list, is 45th on the list of best averages. The main ramifications of using the proposed model is that players from the early years are ranked lower, and spinners are generally ranked higher. The latter is an artefact of the model, in that bowling long spells and taking wickets is now appropriately recognised.

We also see from Table~\ref{TableOfRanks} that seven players have underdispersed data judging from the posterior means of $\nu_i$, verifying that a model capable of handling underdispersed (and overdispersed) counts is required for these data.


\subsection{Model fitting}
\begin{figure}[H]
    \centering
    \includegraphics[width=0.6\textwidth]{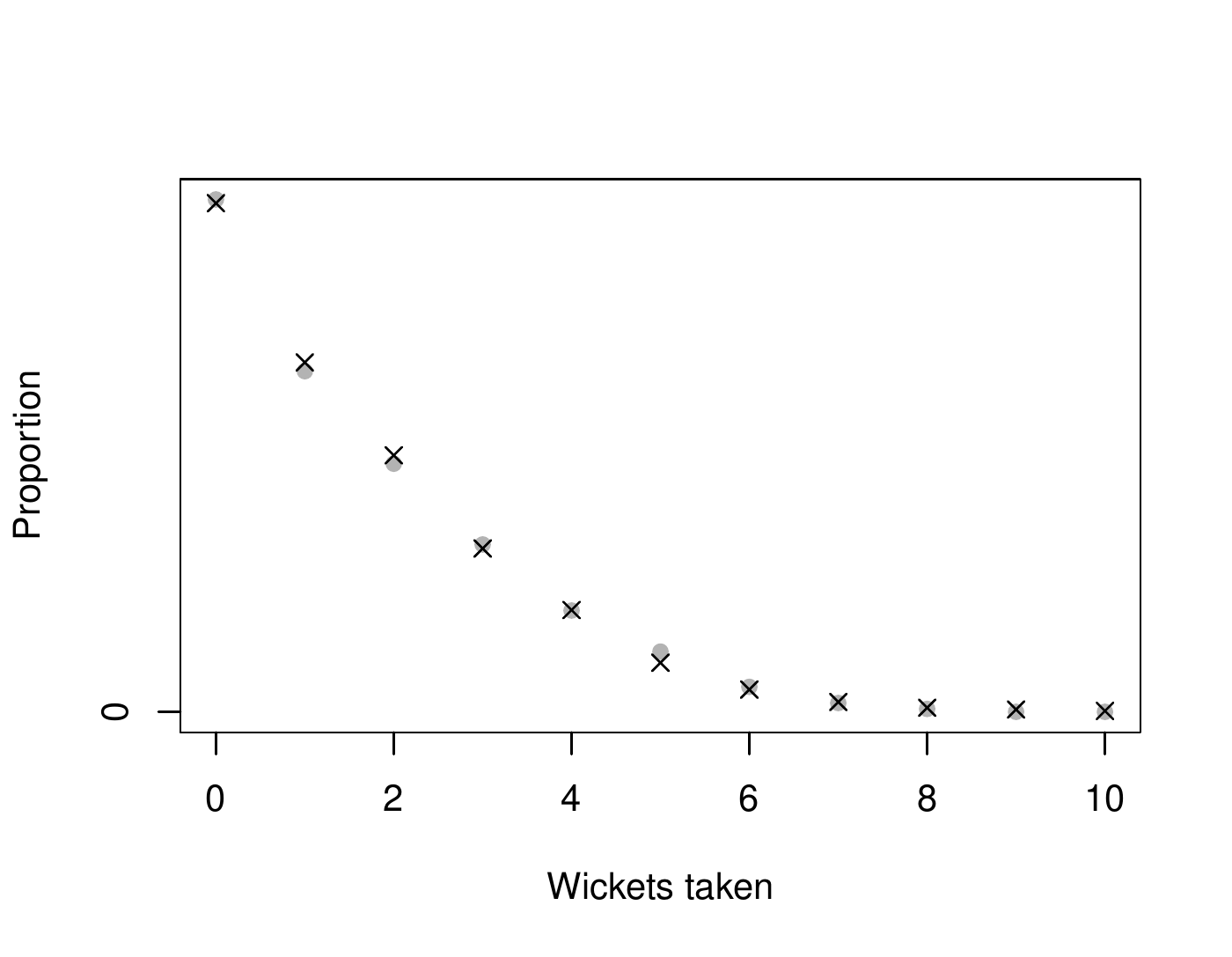}
    \caption{Observed (grey circles) and model-based (crosses) probabilities for each value of wickets}
    \label{OvE_summary}
\end{figure}

The performance of the model is evaluated using the posterior predictive distribution. Namely, due to the small number of observed counts, it is viable to compare the model-based posterior predictive probability for each observed value of 0, 1, \ldots, 10 and compare that to the observed probability in each case. A summary of this information is given in Figure~\ref{OvE_summary}, where we see excellent agreement between the observed and expected proportions at each value of wickets.

\section{Discussion}
The outcome of interest considered here was the number of wickets taken by a bowler in an innings, which was modelled using a truncated mean-parameterised Conway-Maxwell-Poisson distribution. Other classic count models were considered, but it was found that Poisson and negative binomial regression models failed to adequately describe the data. Alternative count models capable of handling both under- and overdispersion may perform equally well, such as those based on the Sichel and Generalised Waring distributions, but alternative models are not pursued further here, principally owing to the good performance of the chosen model.

Additional metrics not formally modelled here are bowling strike rate and bowling economy rate, which concern the number of balls bowled (rather than the number of runs conceded) per wicket and runs conceded per over respectively. Whilst both informative measures, they are typically viewed as secondary and tertiary respectively to bowling average in Test cricket, where time is less constrained than in shorter form cricket. Hence, future work looking at one day international or Twenty20 cricket could consider the triple of average, strike rate and economy rate for bowlers, alongside average and strike rate for batsmen. 

One limitation of this work is that we have ignored possible dependence between players, the taking of wickets can be thought of as a competing resource problem and the success (or lack of) for one player may have an impact on other players on the same team. Indeed, the problem could potentially be recast as one of competing risks. However, this argument is rather circular in that a bowler would not concede many runs if a teammate is taking lots of wickets and this is taken into account through the modelling of the relationship between wickets and runs.

The MPCMP model may have broader use in other sports where there may be bidispersed data, for example modelling goals scored in football matches - stronger and weaker teams are likely to exhibit underdispersion; hockey, ice-hockey and baseball all have small (albeit not truncated) counts as outcomes of interest with bidispersion likely at the team or player level, or both. Moving away from sports to other fields, the MPCMP model could be used to model parity, which is known to vary widely across countries with heavy underdispersion \citep{Barakat2016} and for longitudinal counts with volatile (overdispersed) and stable (underdispersed) profiles at the patient level, where the level of variability may be related to an outcome of interest in a joint modelling setting. Indeed, it will have broader use in any field where counts are subject to bidispersion.

This work has shown that truncated counts subject to bidispersion can be handled in a mean-parameterised CMP model, based on a large dataset, without too much computational overhead. Further methodological work is needed to implement the model in the case of non-truncated counts and to seek faster computational methods.


\newpage
\bibliographystyle{rss}
\bibliography{bowlpaper}

\end{document}